\documentclass[twocolumn,secnumarabic,amssymb, nobibnotes, superscriptaddress,aps, prd]{revtex4-2}
\newcommand{\bea}{\begin{eqnarray}}
\newcommand{\ena}{\end{eqnarray}}
\newcommand{\be}{\begin{equation}}
\newcommand{\ee}{\end{equation}}
\newcommand{\beann}{\begin{eqnarray*}}
\newcommand{\enann}{\end{eqnarray*}}

\usepackage{eurosym}
\usepackage{graphicx}
\usepackage{amsmath}
\usepackage{mathrsfs}
\usepackage{bm}
\usepackage{color}
\usepackage{amsmath}
\usepackage{amssymb}

\usepackage{multirow}
\usepackage{array}
\usepackage{subfig}

\usepackage[utf8]{inputenc}
\usepackage{graphics,graphicx}
\usepackage{amsmath,amssymb,amsfonts,latexsym,cancel}
\usepackage{amsmath,amssymb}
\usepackage[section]{placeins}%force figure to stay in the same section
\usepackage{float}

\begin{document}

\title{Slowly Rotating Neutron Star with Holographic Multiquark Core: I-Love-Q Relations}

\author{Piyabut Burikham}
\email{piyabut@gmail.com}
\affiliation{High Energy Physics Theory Group, Department of Physics,Faculty of Science, Chulalongkorn University, Bangkok 10330, Thailand}

\author{Sitthichai Pinkanjanarod}
\email{quazact@gmail.com, Sitthichai.P@student.chula.ac.th}
\affiliation{High Energy Physics Theory Group, Department of Physics,Faculty of Science, Chulalongkorn University, Bangkok 10330, Thailand}
\affiliation{  Department of Physics, Faculty of Science, Kasetsart University, Bangkok 10900, Thailand}

\author{Supakchai Ponglertsakul}
\email{supakchai.p@gmail.com}
\affiliation{Strong Gravity Group, Department of Physics, Faculty of Science, Silpakorn University, Nakhon Pathom 73000,
	Thailand}

\date{\today}

\begin{abstract}

Moment of inertia ($I$), rotational~(tidal) Love number ($\lambda^{\rm (rot)}$) and quadrupole moment ($Q$) of slowly rotating massive neutron star~(NS) with holographic multiquark~(MQ) core are computed in comparison to pure MQ star. The Chiral Effective Theory~(CET) stiff equation of state~(EoS) is used in the crust of the neutron star.  The dimensionless multipole moments $\bar{I},\bar{\lambda}^{\rm (rot)}, \bar{Q}$ are found to be independent of the rotation parameters and determined completely by the zeroth-order star profile.  Universal ``I-Love-Q'' relations found by Yagi and Yunes~\cite{Yagi:2013bca,Yagi:2013awa} are mostly preserved even in the presence of the MQ core.  Tidal deformation parameter $\bar{\lambda}^{\rm (tid)}$ is also explored in connection with $\bar{I}, \bar{\lambda}^{\rm (rot)}, \bar{Q}$, two kinds of universal I-Love-Q relations are verified.  However, the unique kink in the plots of multipoles with respect to mass and compactness of the population of neutron stars can reveal the existence of massive NS with the MQ core.

\end{abstract}
\maketitle

%\pacs{23.23.+x, 56.65.Dy}

\section{Introduction}\label{sec-introd}

Massive neutron stars with mass around and above two solar masses~($M_{\odot}$) have been observed by a considerable number~\cite{Clark:2002db,Romani:2012rh,Romani:2012jf,vanKerkwijk:2010mt,Linares:2018ppq,Bhalerao:2012xe,Cromartie:2019kug,Nice:2005fi,Demorest:2010bx,Freire:2007sg,Quaintrell:2003pn,Antoniadis:2013pzd,Fonseca:2021wxt,Abbott:2020uma,TheLIGOScientific:2017qsa}.  They naturally require nuclear phase with high density in the core region.  Scan of various EoSs based on the sound speed and adiabatic index suggests that these massive NSs could have quark-matter cores~\cite{Annala:2019puf}.  The nonperturbative nature of strong interaction prevents accurate analysis from the first principles of Quantum Chromodynamics~(QCD) in such extreme situation.  Lattice gauge theory approach has uncertainties from the fermion sign problem when considering high density/chemical potential system.  Bag models assume weakly interacting free quarks within a confinement bag, not necessarily valid for nuclear matter at extreme density presence in the core of massive NS.  

A complementary model inspired by the gauge/gravity duality is to use holographic model of nuclear matter and perform weakly couple calculation in the gravity picture to obtain the physics of strongly interacting gauge theory.  The Sakai-Sugimoto~(SS) model~\cite{ss1,ss2} is a holographic model which shares a number of common features with the QCD.  Variations of the SS model allow a chiral-symmetry-broken deconfined phase~\cite{Aharony_chiral,Bergman:2007wp} with the possibility of the multiquark phase~\cite{bch,Burikham:2011zz}.  

Some of the observed massive NSs have large enough spin for the deformation of the star to be observationally significant.  In such situation, multipole moments of the star become physically important.  For slowly rotating star, the multipole moments can be calculated using perturbative method with respect to the rotation parameters. The moment of inertia, quadrupole moment, and Love number can be used to explore certain aspects of the EoS of the nuclear matter inside a NS.  Yagi et al.~\cite{Yagi:2013bca,Yagi:2013awa} found the ``I-Love-Q relations'' of the multipole moments as well as the unique characterstics of these parameters for slowly rotating hypothetical NS with various EoSs.  The universality of the I-Love-Q relation can be used to test gravitational theory starting from the General Relativity~(GR).  It would be interesting to explore the multipole moments, specifically $I, \lambda, Q$ of the massive NS with potential MQ core with the holographic EoS~\cite{bhp}.  It will be shown that the MQ core generates distinctive multipole moments, $I,\lambda, Q$ from the conventional NS with conventional {\it stiff} nuclear EoS in the Chiral Effective Theory~(CET)~\cite{Tews:2012fj}.  This distinction can be used to distinguish between massive NS with MQ and NS with conventional nuclear core.  

This work is organized as the following.  Section~\ref{sec-eos} describes the EoS we use for the MQ and the nuclear matter, more details are given in \cite{Pinkanjanarod:2020mgi}.  Section~\ref{secgam} presents the spacetime metric and stress tensor of the slowly rotating star in GR up to the second order in the rotation parameters. Section~\ref{sec-eom} discusses the equations of motion at the zeroth, first and second order of the perturbations with respect to the rotation.  Numerical results are presented in Section~\ref{sec-num} and Section~\ref{sec-con} concludes our work.

\section{EoS for massive neutron star with holographic mulitquark core }\label{sec-eos}

The interior composition of massive NS can be studied via observations and various theoretical models of hadronic matter in a cold environment. At low density, quarks are strongly coupled and confined within hadrons. Generally, the dynamical behaviour of hadrons could be quantitatively described by mesons exchange based on CET as a low-energy effective theory since their interactions are weak and short-range. Many important parameters in CET can be calculated by using the perturbative power expansion in terms of pion mass $M_\pi$ and the chiral symmetry breaking scale $\Lambda_\chi \sim 1$ GeV \cite{Machleidt:2011zz}.  The EoS for the cold nuclear matter could be obtained by considering two-nucleon and three-nucleon interactions, for greater accuracy, within the framework of CET \cite{Hebeler:2013nza}. However, beyond the nuclear saturation density $n_0 \approx 0.16 $ fm$^{-3}$, there are uncertainties associated with a series of polytropic EoS determined from the extended CET that leads to 3 possibilities, i.e., the soft, the intermediate, and the stiff EoS~\cite{Hebeler:2013nza}. 

In the core of massive NS where the density is extremely large, the quarks are expected to be effectively deconfined since they can hardly distinguish one baryon from the neighbouring ones. Dripping quarks from one baryon become closer to the others, resulting in unclear boundaries of the original baryonic bound state. Holographic QCD such as the variations of SS model suggests \cite{Aharony_chiral,Bergman:2007wp,bch} that these deconfined quarks could form MQ bound states while chiral symmetry is still broken. Additionally, holographic MQ stars, which are assumed to be entirely in holographic MQ state, were studied in \cite{bhp} where the preliminary estimates of the mass of NS with MQ core could be as high as $3M_{\odot}$.

It was found in \cite{Pinkanjanarod:2020mgi} that the core of massive NS could be in the holographic MQ phase which is more thermodynamically prefered than the {\it stiff} CET nuclear matter for a certain range of the model parameters. In this work, we extend our analysis to include effects of slow rotation by considering the massive neutron stars with MQ core obeying EoS from the holographic SS model.  The nuclear crust is assumed to obey stiff EoS from the CET following \cite{Pinkanjanarod:2020mgi}. 

\subsection{Equation of state of the multiquark core}  \label{sec-eosmq}

According to the holographic model of multiquark proposed and studied in \cite{bch,bhp}, the EoS of the holographic multiquark depends on number density $n$ and a relative number of colour charges per mulitquark $n_s$. At large $n$, pressure $P$ and density $\rho$ of high-density multiquark (mqh) are given, in the dimensionless form, by
\bea
P &=& k n^{7/5},  \notag \\
\rho c^2 &=& \rho_{c} c^2+\frac{5}{2}P_c+\mu_{c}\left(n-n_{c}\right) \notag \\
&&+kn_{c}^{7/5}-\frac{7k}{2}n_{c}^{2/5},  \label{eosmqh}
\ena
where $n_c$ is a critical number density at the transition between mqh and low-density multiquark (mql) while $P_c = P(n_c)$, $\rho_c = \rho(n_c)$, and $\mu(n_c) = \mu_c$ are pressure, density, and mutiquark chemical potential energy at the transition. For $n_{s}=0: n_{c}=0.215443, \mu_{c}=0.564374$ whereas for $n_{s}=0.3: n_{c}=0.086666, \mu_{c}=0.490069$, while $k=10^{-0.4}$ for both cases.
At smaller $n$, EoS of mql are given by
\bea
P &=& a n^{2}+b n^{4},  \notag \\
\rho c^2 &=& \mu_{0}n+a n^{2}+\frac{b}{3}n^{4},  \label{eosmql}
\ena
where the onset chemical potential of the multiquark phase $\mu_0 =\mu(n=0)$.  For $n_{s}=0$, $a=1, b=0, \mu_{0}=0.17495$ while for $n_{s}=0.3$, $ a=0.375, b=180.0, \mu_{0}=0.32767$. In Ref.~\cite{Pinkanjanarod:2020mgi}, it has been shown that only MQ with $n_{s}=0.3$ can interpolate well between the CET EoS at low densities and perturbative QCD~(pQCD) at much higher energy densities,
therefore we only consider $n_{s}=0.3$ in this work.

Note that parameters represented in (\ref{eosmqh}) and (\ref{eosmql}) are all in dimensionless form. Conversions of thermodynamical quantities from dimensionless to conventional physical units depend only on the energy density scale $\epsilon_s$ expressed in GeV~fm$^{-3}$, defined in \cite{bhp,Pinkanjanarod:2020mgi,Pinkanjanarod:2021qto}. The pressure and mass density scale with $\epsilon_s$ as $P, \rho \sim \epsilon_s$.  The mass and radius of the pure multiquark star have the same scaling $M, R \sim \epsilon_s^{-1/2}$ while the compactness $M/R$ is independent of $\epsilon_s$.

\subsection{Equation of state of the nuclear matter crust}  \label{sec-eosnuc}

As described in details in \cite{Pinkanjanarod:2020mgi}, EoS for nuclear matter in the NS could be divided into 3 regions.  For very low densities, EoS of degenerate nucleons can be found in Table 7 of Ref. \cite{Hebeler:2013nza}.  For intermediate densities, it is approximated by a series of polytropes as shown in Eq.~(18) of Ref.~\cite{Pinkanjanarod:2020mgi}. Then at slightly higher densities ranging from $75.1$ MeV fm$^{-3}$ to $165.3$ MeV fm$^{-3}$, EoS of weakly interacting nucleons consisting of chiral quarks described by CET can be found in Eqs. $(19)$ and $(20)$ of Ref. \cite{Pinkanjanarod:2020mgi}. For nuclear matter beyond a typical density $\rho_1 = 165.3$ MeV fm$^{-3}$ up to the transition density, EoS could be obtained from an extension of CET based on the nucleon-nucleon and three-nucleon interactions using asymmetric nuclear matter as expressed in Table $5-6$ of Ref.~\cite{Hebeler:2013nza} and could also be described by a set of polytropes as expressed in Eq.~(21) of Ref.~\cite{Pinkanjanarod:2020mgi}. This results in 3 possible extensions: soft, intermediate, and stiff extended CET EoS.

Furthermore, a phase transition between multiquark state and extended CET nuclear matter has been studied in \cite{Pinkanjanarod:2020mgi}. It was found by studying the pressure v.s. quark chemical potential or $P-\mu$ diagram that there are possible transitions only from multiquark state to stiff extended CET nuclear matter with sensible energy density scales $\epsilon_s$ ranging from $26 - 28$ GeV fm$^{-3}$. Additionally, we found the multiquark state is preferred over stiff extended CET nuclear matter at densities higher than the transition density.

\section{Background metric and energy momentum tensor}  \label{secgam}

A uniformly rotating neutron star (NS) can be perturbatively described by a slow-rotation expansion in an isolated non-rotating background solution. Such a neutron star solution can be expressed in Boyer-Lindquist coordinates as~\cite{Yagi:2013awa}
\begin{align}
ds^2 &= -e^{\bar{\nu}_0(r)}\left[1+2\epsilon^2 \bar{H}_2(r)P_2(\cos\theta)\right]dt^2 \nonumber \\
& + e^{\bar{\lambda}_0(r)}\left[1 + \frac{2\epsilon^2 \bar{S}_2(r) P_2(\cos\theta)}{r-2\bar{m}(r)}\right]dr^2  \nonumber \\
& + r^2\left[1+2\epsilon^2 \bar{K}_2(r)P_2(\cos\theta)\right] \nonumber \\
& \times \left(d\theta^2 + \sin^2\theta \left[ d\phi -\epsilon\omega(r,\theta)dt \right]^2
\right) + \mathcal{O}(\epsilon^3), \label{metric}
\end{align}
where $\bar{m}(r)$ is often related by
\begin{align}
e^{\bar{\lambda}_0(r)} &= \left(1-\frac{2\bar{m}(r)}{r}\right)^{-1}.
\end{align}
This $\bar{m}(r)$ can be interpreted as an accumulated mass function. At the NS's surface $r=R$ and $\bar{m}(R)=M$ where $M$ is a total mass of the star. We thus have
\begin{align}
e^{\bar{\nu}_0(R)} = e^{-\bar{\lambda}_0(R)} = 1-\frac{2M}{R}. \label{exterior}
\end{align}
The expansion parameter $\epsilon$ denotes the order of approximation. At the first order in $\epsilon$, neutron star's rotation is introduced by the angular velocity \cite{Hartle:1967he}
\begin{align}
\omega(r,\theta) &= \Omega- \bar{\omega}_1(r)\left(-\frac{1}{\sin\theta}\frac{dP_1}{d\theta}\right), \label{Omeq}
\end{align}   
where $P_\ell(\cos(\theta))$ is the $\ell$-th order Legendre polynomial. The second term on the right-hand side of (\ref{Omeq}), given by $\Omega-\omega$, is the angular velocity of star content at $(r,\theta)$ seen by the free falling observer. At the second order in $\epsilon$, the deformations of NS are denoted by second-order quantities $\bar{H}_2(r),\bar{S}_2(r),\bar{K}_2(r)$.

Here we are studying rotational effect of neutron star perturbatively. This perturbative approach is simply valid when the differences between physical quantities in rotating and nonrotating case are small~\cite{Yagi:2013mbt}. If we follow standard polar coordinate ($r,\theta$), there will be some point where this perturbation technique is no longer valid. For instance, the pressure $P$ of nonrotating star vanishes at the surface (at some value of $r=R$), since the shape of the star changes when it rotates, thus the pressure is non-zero in this case. Therefore the perturbation scheme based on a ratio of density ($\frac{\Delta\rho}{\rho}$ ) becomes infinitely large and invalid. To overcome this issue, Hartle	~\cite{Hartle:1967he} introduced a coordinate transformation ($r,\theta$)$\to$($\bar{r},\Theta$), it is given by
\begin{align}
\rho\left[r(\bar{r},\Theta),\Theta\right] = \rho(\bar{r}),~~~~~~~\Theta=\theta.
\end{align}
The radial coordinate $\bar{r}$ is chosen such that $\rho$ and $P$ are the same in both rotating and nonrotating configurations. The radial coordinate $\bar{r}$ is expanded by
\begin{align}
r(\bar{r},\Theta) = \bar{r} + \epsilon^2\xi_2(\bar{r})P_2(\cos\Theta).
\end{align}
Here and henceforth, any metric coefficients expressed without ``bar'' means they are written in $\bar{r}$ coordinate e.g. $H_2(\bar{r}) \equiv \bar{H}_2(r)$. We also denote derivative with respect to $\bar{r}$ with $'$ here and henceforth.

A matter content inside uniformly rotating NS will be modeled by perfect fluid material. The stress-energy momentum tensor is defined by
\begin{align}
T_{\mu\nu} &= (\rho+P)u_\mu u_\nu + Pg_{\mu\nu}, \label{energytensor}
\end{align}
where four-velocity of the perfect fluid is normalized by $u_\mu u^{\mu}=-1$. This is given by
\begin{align}
u^\mu &= \left(u^0,0,0,\epsilon\Omega u^0\right),
\end{align} 
where the time component of four-velocity is obtained from~\cite{Hartle:1968si}
\begin{align}
u^0 &= \left[-\left(g_{00} + 2\epsilon\Omega g_{03} + \epsilon^2\Omega^2 g_{33}\right)\right]^{-1/2}, \nonumber \\
&= e^{-\nu/2}\left[1 + \epsilon^2\frac{e^{-\nu}}{2}\left\{ (\bar{r} \bar{\omega}_1 \sin\theta)^2 - e^{\nu}\left(2H_2+\nu'\xi_2\right)P_2  \right\} \right].
\end{align}
The EoSs to be used is the holographic MQ EoS in the core connecting with the stiff CET EoS in the crust of the NS.  Pure MQ star is also considered for comparison. 

\section{Equation of motion} \label{sec-eom}

In this section, we will construct differential equations corresponding to $i)$ isolated, nonrotating neutron star, $ii)$ slowly rotating neutron star to linear order in spin $iii)$ slowly rotating neutron star to quadratic order in spin. Then we discuss interior and exterior solutions of these equations. By matching both solutions at the boundary i.e., at the surface of NS, we obtain useful physical quantities of the star such as, total mass, radius, moment of inertia, quadratic moment and rotational Love number. 

This part is a review of analyses in \cite{Hartle:1967he,Yagi:2013mbt} where we elaborate more on the independence of $I, \lambda, \bar{Q}$ to the rotation parameters as well as the basic formalism on which we analyze the rotational properties of NS with MQ core and pure MQ stars. 

\subsection{Einstein equations $\mathcal{O}(\epsilon^0)$}

From the metric (\ref{metric}) and stress-energy tensor (\ref{energytensor}), the ($t,t$) and $(\bar{r},\bar{r})$ components of the Einstein field equations are given by
\begin{align}
m' &= 4\pi \bar{r}^2\rho,\label{TOV1} \\ 
\nu' &= 2\left(\frac{4\pi \bar{r}^3 P+m}{\bar{r}\left(\bar{r} -2 m\right)}\right). \label{TOV2}
\end{align}
The Tolman-Oppenheimer-Volkov (TOV) equation can be obtained from the radial component of conservation of energy i.e., $\nabla_{\mu} T^{\mu \bar{r}} = 0$,
\begin{align}
P' &= -\left(\frac{4\pi \bar{r}^3 P+m}{\bar{r}\left(\bar{r} -2 m\right)}\right)\left(\rho+P\right). \label{TOV3}
\end{align}
With equation of state given by (\ref{eosmqh}), (\ref{eosmql}), equations (\ref{TOV1}--\ref{TOV3}) form a system of coupled first order ordinary differential equations. These equations can be solved numerically when appropriate boundary conditions are specified. Outside the star, $\rho=0,~P=0$, these equations admit the Schwarzschild solution (\ref{exterior}) with mass $M$ and $\bar{r}>R$. 

An extra attention must be taken when considering the initial condition of $\nu(\bar{r}_{0})=\nu_{c}$. Since our field equations are shift-invariant in $\nu$, therefore adding some constant to $\nu$ does not change the whole equation of motion. Consequently, $\nu_{c}$ at the center must be chosen so that \cite{Yagi:2013mbt}
\begin{align}
e^{\nu(R)} &= 1-\frac{2M}{R}, \label{nuceq}
\end{align}
at the star surface.  In practice, we set the cutoff radius $\bar{r}_{\rm min}=10^{-6}$ and integrate outward until we reach the star surface whereas $P(R)=0$ and $m(R)=M$. The resulting zeroth-order star profiles are then used to calculate the first order perturbation $\bar{\omega}_{1}$. The choice of $\nu$ satisfying (\ref{nuceq}) is necessary in the correct calculation of the first and second order perturbations.

\subsection{Equation in linear order $\mathcal{O}(\epsilon^1)$}
At the linear level, the only non-vanishing component of Einstein field equation is ($t,\phi$). This yields
\begin{align}
0 &= \frac{d^2\omega_1}{d\bar{r}^2} + 4\left[\frac{1-\pi \bar{r}^2\left(\rho+P\right)e^{\lambda}}{\bar{r}}\right]\frac{d\omega_1}{d\bar{r}} \nonumber \\
&~~~- 16\pi\left(\rho+P\right)e^{\lambda}\omega_1, \label{1storderequation}
\end{align}
where $\omega_{1}(\bar{r})\equiv \bar{\omega}_{1}(r)$ and
\begin{align}
e^{-\lambda(\bar{r})} &=\left( 1-\frac{2m(\bar{r})}{\bar{r}}\right). \label{lamb1}
\end{align}
To solve this equation, one needs to explore asymptotic behavior of $\omega_1$ at the centre of the star and the exterior. Outside the star, there is no matter i.e. $\rho=P=0$ and $m(\bar{r})=M$. In this region, (\ref{1storderequation}) becomes exactly solvable and its solution is given by \cite{Yagi:2013awa,Hartle:1967he}
\begin{align}
\omega_1^{\text{out}} &= \Omega\left(1-\frac{2I}{\bar{r}^3}\right), \label{exterior1}
\end{align}
where the moment of inertia is defined by $I\equiv S/\Omega$. Two constants $S$ and $\Omega$ can be interpreted as spin angular momentum and angular velocity of the star, respectively. The linearity of (\ref{1storderequation}) and the asymptotic relation $\omega_{1}(\bar{r}\to\infty)=\Omega$ implies that $\omega_{1}$ must scale with $\Omega$.  Dividing (\ref{1storderequation}) by $\Omega$ and solve with the scaled boundary condition $\omega_{1}(\bar{r}\to\infty)/\Omega =1$ results in the scaled inner and outer solutions which are independent of $\Omega$.  Consequently, the outer scaled solution $\omega_{1}^{\rm out}/\Omega$ must also be independent of $\Omega$, and $I$ given in (\ref{lamb1}) is automatically independent of the rotation parameters $\omega_{1}^{\rm out}, \Omega$.  Remarkably, $I$ represents {\it intrinsic} properties of star with respect to slow rotation and it depends only on the zeroth-order star profile. 

For interior solution, we perform Taylor expansion on (\ref{1storderequation}) around the star centre. The function $\omega_1$ behaves regularly as
\begin{align}
\omega_1^{\text{in}} &= \omega_c + \frac{8\pi}{5}\left(\rho_c+P_c\right)\omega_c \bar{r}^2 + \mathcal{O}(\bar{r}^3). \label{BCin4}
\end{align}
From the zeroth order, complete profiles of $m,\rho$ and $P$ are obtained. Then (\ref{1storderequation}) can be numerically integrated starting from (\ref{BCin4}) until we reach the surface $\bar{r}=R$. With some test values of $\Omega,$ and $\omega_c$, $I$ can be determined from continuity of $\omega_1$ i.e.,
\begin{align}
\omega_1^{\text{in}}(R) &= \omega_1^{\text{out}}(R),~~~~~~~~\frac{d}{d\bar{r}}\omega_1^{\text{in}}(R) = \frac{d}{d\bar{r}}\omega_1^{\text{out}}(R). \label{matchcond}
\end{align}
Alternatively, the moment of inertia can be obtained via \cite{Yagi:2013awa,Hartle:1967he}
\begin{align}
I &= \frac{8\pi}{3\Omega}\int_{0}^{R} \frac{\bar{r}^5\left(\rho+P\right)e^{-\left(\nu+\lambda\right)/2}}{\bar{r} - 2m(\bar{r}) } \omega_1 d\bar{r},
\end{align}
provided that (\ref{nuceq}) is satisfied.  We have checked that the moment of inertia calculated via the matching method and the formula above are in perfect agreement. The numerical results also confirm the independence of $I$ to rotation parameters.  For demonstration purpose, it is convenient to define dimensionless moment of inertia
\begin{align}
\bar{I} &\equiv \frac{I}{M^3}.
\end{align}

\subsection{Equation in quadratic order $\mathcal{O}(\epsilon^2)$}

At quadratic order in spin, the equations of motion involve only $H_2,K_2,S_2$ and $\xi_2$. In fact, it turns out that there are two evolution equations and two constraints. From energy conservation, $\theta$ is the only non-zero component $\nabla_{\mu}T^{\mu\theta}=0$. This gives
\begin{align}
\xi_2(\bar{r}) &= -\frac{e^{-\nu}\bar{r}\left(\bar{r}-2m\right)\left(\bar{r}^2\omega_1^2 + 3e^{\nu} H_2\right)}{3\left(m+4\pi P \bar{r}^3\right)}.
\end{align}
Non-vanishing components of Einstein field equations at quadratic order are \\
$(\theta,\theta)-(\phi,\phi):$
\begin{align}
S_2(\bar{r}) &= -\left(\bar{r}-2m\right)H_2 + \frac{1}{6}e^{-\lambda-\nu}\bar{r}^4\left[\left(\bar{r}-2m\right)\omega'^2_1 \right. \nonumber \\ 
&~~~\left. + 16\pi \bar{r}\left(\rho+P\right)\omega_1^2  \right], \label{Seq}
\end{align}
$(\bar{r},\theta):$
\begin{align}
K'_2 &= -H'_2 + \left[\frac{\bar{r}\left(1-4\pi P \bar{r}^2\right)-3m}{\bar{r}\left(\bar{r}-2m\right)}\right]H_2 \nonumber \\
&~~~+ \left[\frac{\bar{r}\left(1+4\pi P \bar{r}^2\right)-m}{\bar{r}\left(\bar{r}-2m\right)^2}\right]S_2, \label{Keq}
\end{align}
$(\bar{r},\bar{r}):$
\begin{align}
H'_2 &= \left[\frac{m-\bar{r}\left(1+4\pi P \bar{r}^2\right)}{\left(\bar{r}-2m\right)}\right]K'_2 + \left[\frac{2}{\bar{r}-2m} \right]K_2 \nonumber \\
&~~~+ \left[\frac{3-4\pi \bar{r}^2\left(\rho+P\right)}{\bar{r}-2m}\right]H_2 + \left[\frac{1+8\pi P \bar{r}^2}{\left(\bar{r} - 2m\right)^2}\right]S_2 \nonumber \\
&~~~+ \frac{\bar{r}^3}{12}e^{-\nu}\omega_1'^2 + \frac{4\pi}{3}\frac{\left(\rho+P\right)\bar{r}^4e^{-\nu}}{\left(\bar{r}-2m\right)}\omega_1^2. \label{Heq}
\end{align}
Note that we can replace $S_2$ in (\ref{Keq}) and (\ref{Heq}) with (\ref{Seq}). In the exterior region where $\lambda,\nu$ and $\omega_1$ can be expressed as (\ref{exterior}) and (\ref{exterior1}), the solutions of evolution equation $H'_2,K'_2$ can be written as \cite{Yagi:2013awa} 
\begin{align}
H_2^{\rm out} &= \frac{1}{\bar{r}^4}\left(1+\frac{1}{\mathcal{C}}\right)\left(I \Omega\right)^2 + A\left[\mathcal{C}-\frac{3}{\mathcal{C}}+\frac{1}{2-4\mathcal{C}}+\frac{5}{2} \right. \nonumber \\
&~~~+ \frac{3\left(2\mathcal{C}-1\right)}{2\mathcal{C}^2}\ln\left(1-2\mathcal{C}\right) \left. \right], \\
K_2^{\rm out} &= -\frac{1}{\bar{r}^4}\left(2+\frac{1}{\mathcal{C}}\right)\left(I \Omega\right)^2 + \frac{3A}{\mathcal{C}}\left[1+\mathcal{C}-\frac{2\mathcal{C}^2}{3} \right. \nonumber \\ 
&~~~+ \frac{\left(1-2\mathcal{C}^2\right)}{2\mathcal{C}}\ln\left(1-2\mathcal{C}\right) \left. \right],
\end{align}
where $\mathcal{C}\equiv M/\bar{r}$ and $A$ is integration constant to be determined later.  It is also useful to define star's compactness as $\mathcal{C}\rvert_{R}\equiv C $. For the interior solutions, expanding (\ref{Keq}--\ref{Heq}) around $\bar{r}=0$ yields
\begin{align}
H_2 &= B \bar{r}^2 + \mathcal{O}(\bar{r}^4), \\
K_2 &= -B \bar{r}^2 + \mathcal{O}(\bar{r}^4),
\end{align}
where $B$ is arbitrary constant. These constants $A$ and $B$ will be determined by matching the boundary conditions at the surface of the star,
\begin{align}
H_2^{\rm in}(R) = H_2^{\rm out}(R),~~~~~~K_2^{\rm in}(R) = K_2^{\rm out}(R). \label{matchcond1}
\end{align}
In principle, one can numerically integrate (\ref{Keq}--\ref{Heq}) starting from the initial conditions inside the star until the matching conditions above are satisfied. In practice, we adopt Hartle's approach \cite{Hartle:1967he} for solving this system of equation. First, we write down interior solution as a sum of particular solution ($H_2^{p}$) and the product of an arbitrary constant ($C_1$) and the homogeneous solution ($H_2^{h}$)
\begin{align}
H_2^{\rm in} &= H_2^{p} + C_1 H_2^{h}, \\
K_2^{\rm in} &= K_2^{p} + C_1 K_2^{h}.
\end{align}
For a given value of $B$, the particular and homogeneous solutions can be computed numerically. The unknown constants $A$ and $C_1$, will be then determined from simple algebraic equations (\ref{matchcond1}). The quadrupole moment can be calculated from~\cite{Hartle:1967he,Hartle:1968si,Yagi:2013awa}
\begin{align}
Q^{({\rm rot})} = -\frac{\left(I\Omega\right)^2}{M} - \frac{8}{5}AM^3. \label{Qeq}
\end{align}
The dimensionless spin-induced quadrupole moment can be defined as
\begin{align}
\bar{Q} &\equiv -\frac{Q^{({\rm rot})}M}{\left(I\Omega\right)^2}, \nonumber \\
&=1+\frac{8A}{5}\left(\frac{M^2}{I\Omega}\right)^2.
\end{align}
In the outer region, since $P$ and $\rho$ are zero, the terms proportional to $\omega_1^2$ in (\ref{Seq}), (\ref{Keq}), and (\ref{Heq}) become zero.  Using (\ref{exterior1}), the terms involving ${\omega'_{1}}^{2}$ are thus proportional to $S^{2}$. Dividing (\ref{Seq}--\ref{Heq}) with $S^{2}$ throughout results in the rescaling of $S_{2}, K_{2}, H_{2}$ with $S^{2}$ and the equations of motion which are independent of $S$. So we can conclude that all solutions of $S_{2}, K_{2}, H_{2}$ in the outer region must simply scale with $S^{2}$, i.e., $A\sim S^{2}$~(and the scaled solutions $(S_{2}, K_{2}, H_{2})/S^{2}$ are independent of $S$).  As a consequence, $Q^{(\text{rot})}$ given by (\ref{Qeq}) will scale with $S^{2}$ and $\bar{Q}$ is always independent of $S, \omega, \Omega$. This {\it universality} has also been verified by our numerical results.  For a given star profile with small but arbitrary $\Omega, \omega$, the values of $I$ and $\bar{Q}$ are independent of the rotation parameters.  

Since now we have isolated the star profiles upto the second order in spin, we can define $\ell=2$ rotational Love number as
\begin{align}
\lambda^{\rm (rot)} &\equiv - \frac{Q^{\rm (rot)}}{\mathcal{E}^{\rm(rot)}}, \nonumber \\
&=-\frac{Q^{\rm (rot)}}{\Omega^{2}},
\end{align}
where $\mathcal{E}^{\rm (rot)}$ is the quadrupolar contribution of the centrifugal potential \cite{Yagi:2013awa}. In addition, the quantity $\mathcal{E}^{\rm (rot)}$ can be expressed as $\Omega^2$ in the Newtonian limit \cite{Mora:2003wt}. The rotational Love number measures how much neutron/MQ star deforms away from spherical shape due to its spin. The dimensionless rotational Love number is given by
\begin{align}
\bar{\lambda}^{\rm (rot)} &\equiv \frac{\lambda^{\rm (rot)}}{M^5}.
\end{align}
Thus one can relate the star's moment of inertial, quadrupole moment and rotational Love number as
\begin{align}
\bar{\lambda}^{\rm (rot)} &= {\bar{I}}^2\bar{Q}.
\end{align}
From (\ref{Qeq}), it is obvious that both $\lambda^{\rm (rot)}, \bar{\lambda}^{\rm (rot)}$ are independent of rotation parameters and determined only by the zeroth-order star profile.

\section{Numerical results}   \label{sec-num}

\begin{figure}[H]
	\centering
	\includegraphics[width=0.48\textwidth]{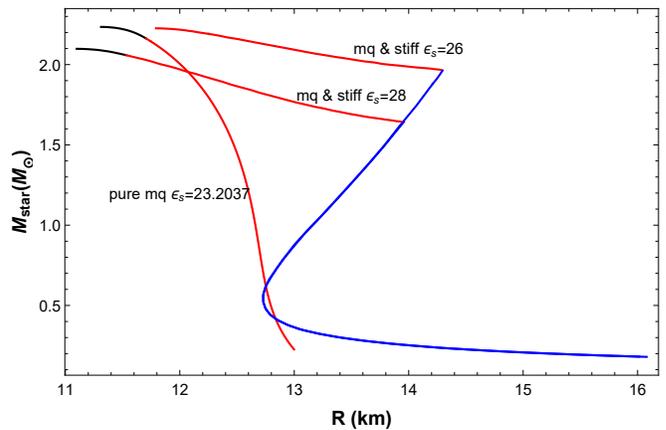}
	\caption{$M-R$ diagram of NS with MQ core and pure MQ star.  Energy density scale $\epsilon_{s}$ is in GeVfm$^{-3}$ unit. }
	\label{fig:MR}
\end{figure}

\begin{figure}[H]
	\centering
	\includegraphics[width=0.48\textwidth]{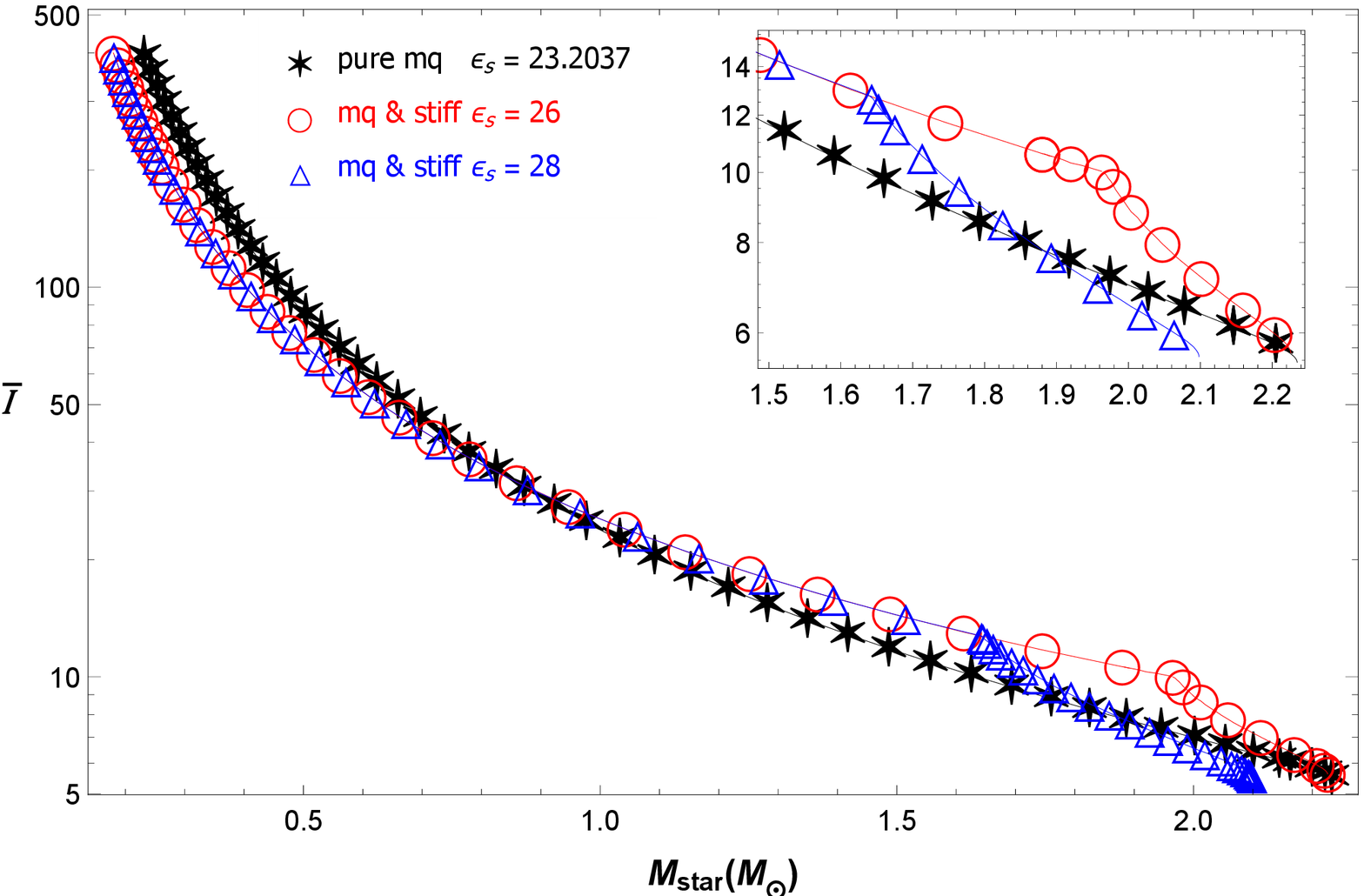}
	\includegraphics[width=0.48\textwidth]{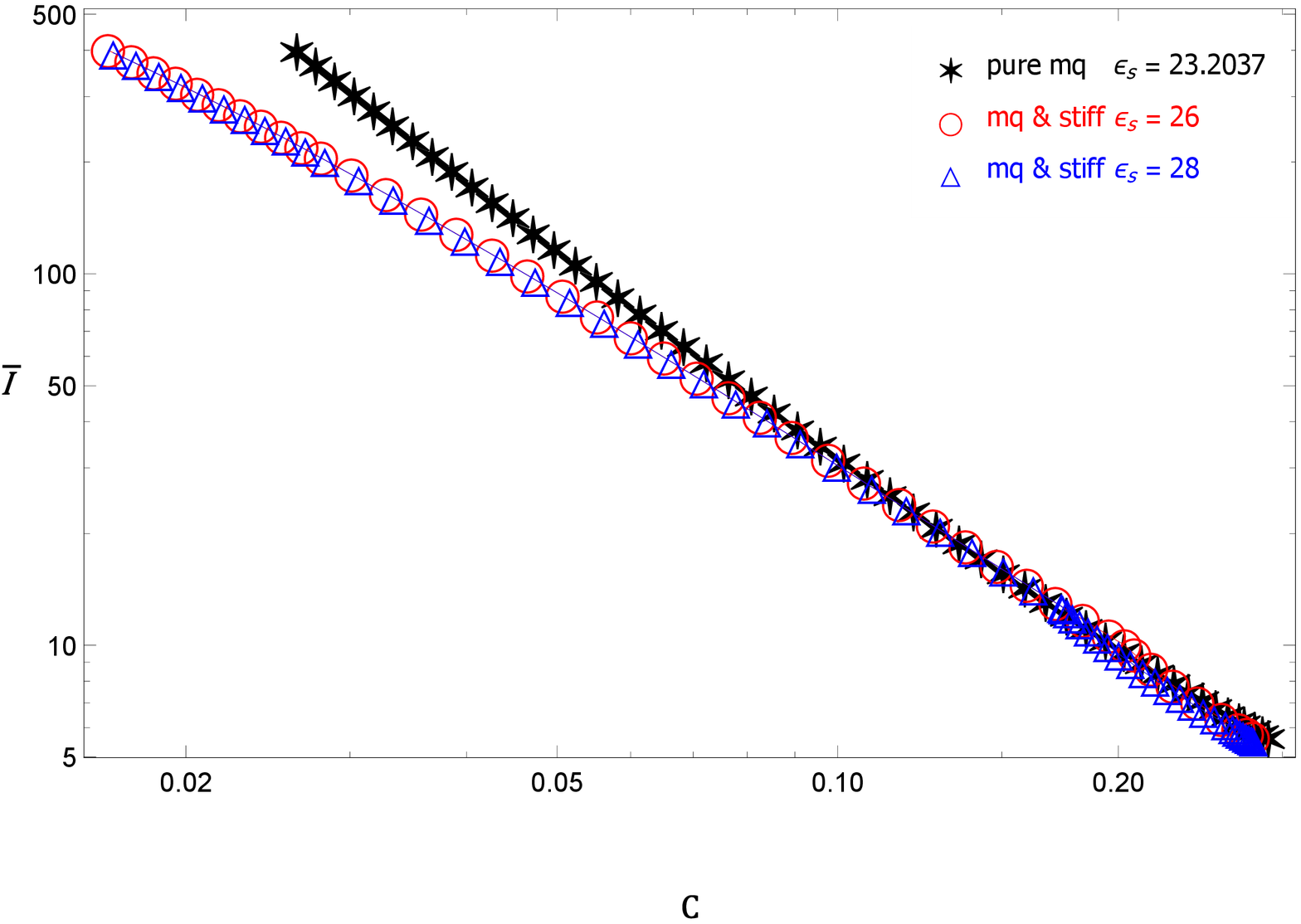}
	\caption{Upper: Dimensionless moment of inertia parameter $\bar{I}$ vs. mass of the star $M$. Lower: $\bar{I}$ vs. compactness $C$}
	\label{fig:IMC}
\end{figure}

\begin{figure}[H]
	\centering
	\includegraphics[width=0.44\textwidth]{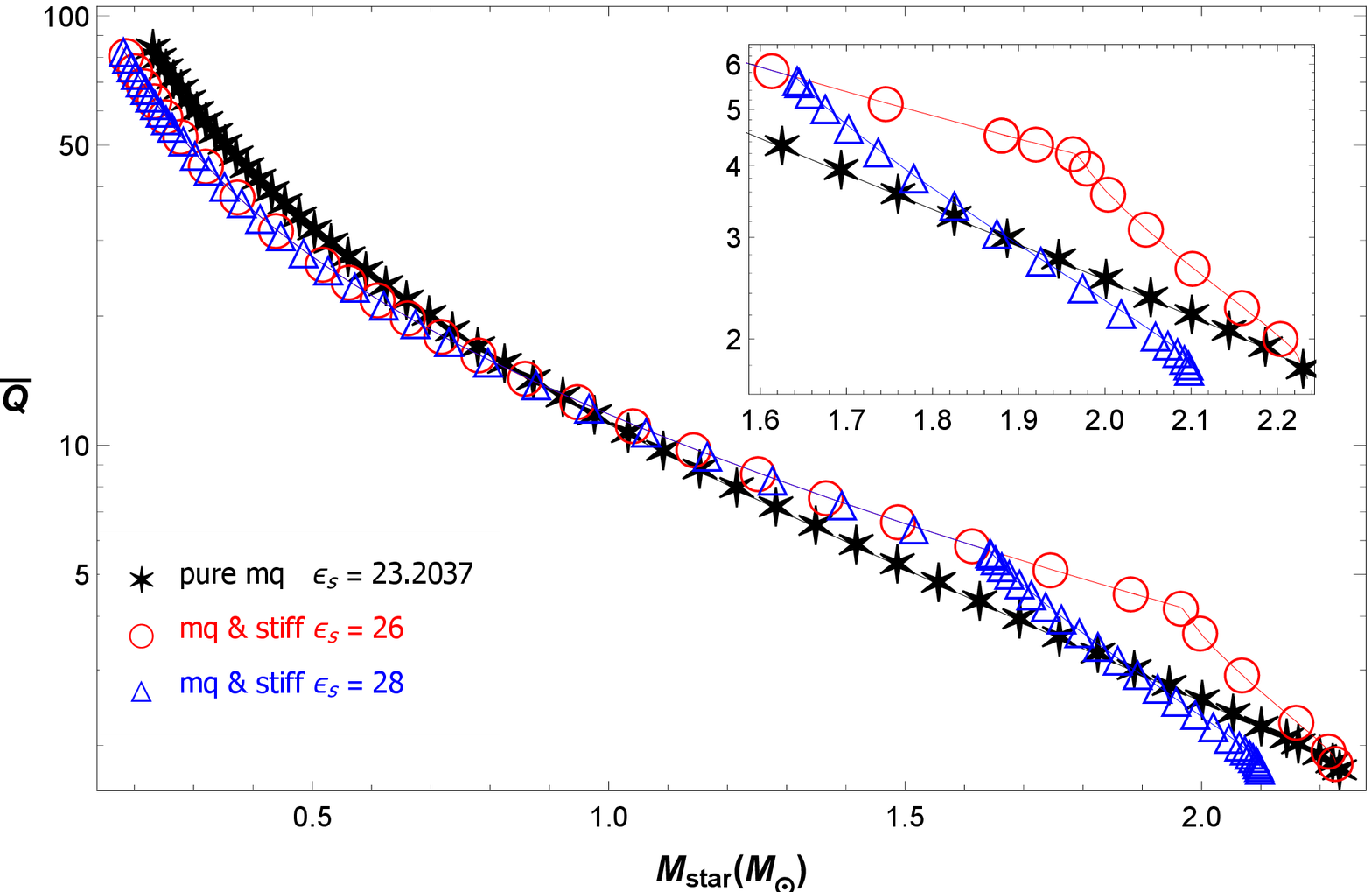}
	\includegraphics[width=0.44\textwidth]{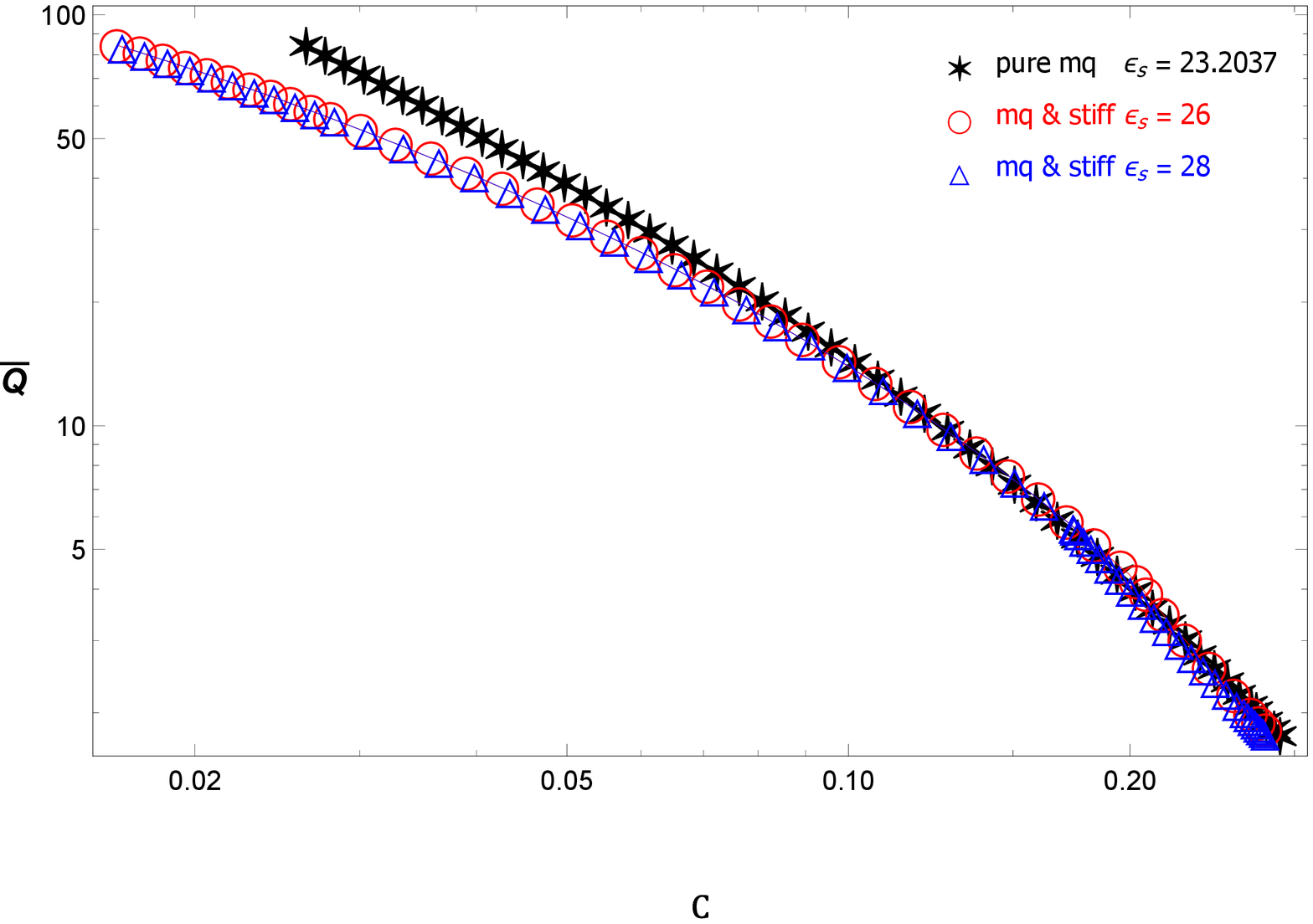}
	\caption{Upper: Dimensionless quadrupole moment parameter $\bar{Q}$ vs. $M$. Lower: $\bar{Q}$ vs. $C$ }
	\label{fig:QMC}
\end{figure}

\begin{figure}[H]
	\centering
	\includegraphics[width=0.44\textwidth]{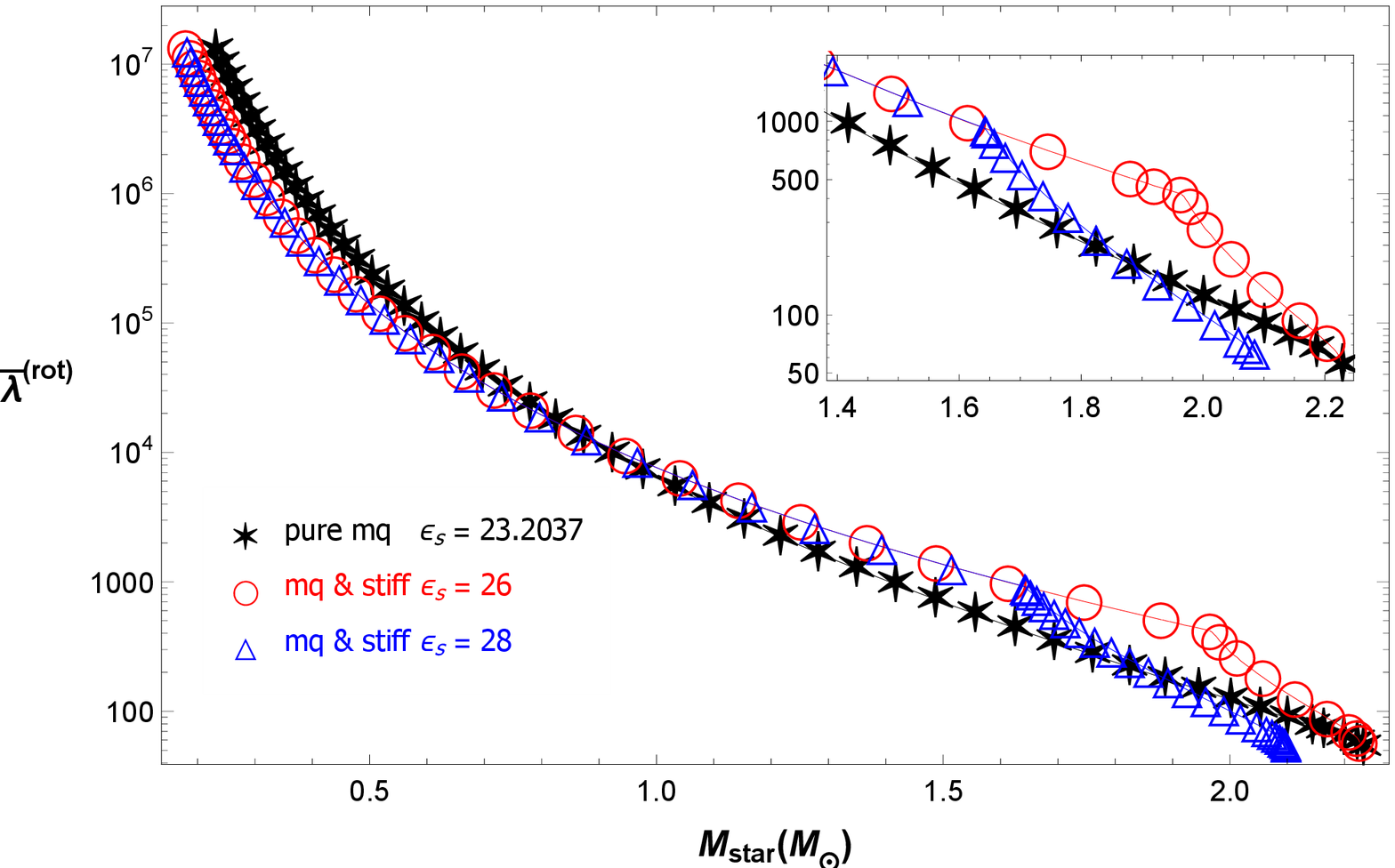}
	\includegraphics[width=0.44\textwidth]{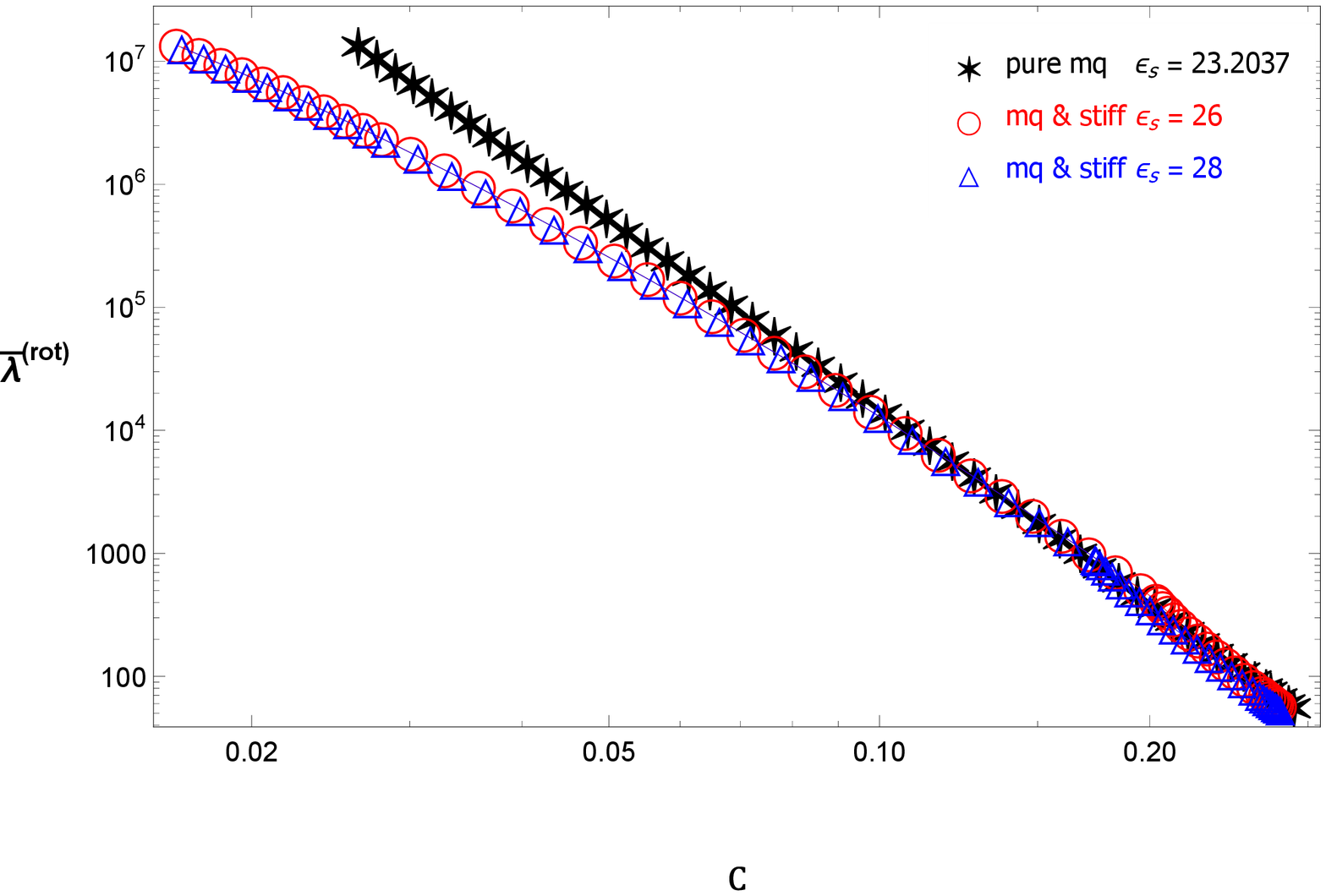}
	\caption{Upper: Rotational Love number $\bar{\lambda}^{\rm (rot)}$ vs. mass of the star $M$. Lower: $\bar{\lambda}^{\rm (rot)}$ vs. compactness $C$ }
	\label{fig:LMC}
\end{figure}

\begin{figure}[H]
	\centering
	\includegraphics[width=0.48\textwidth]{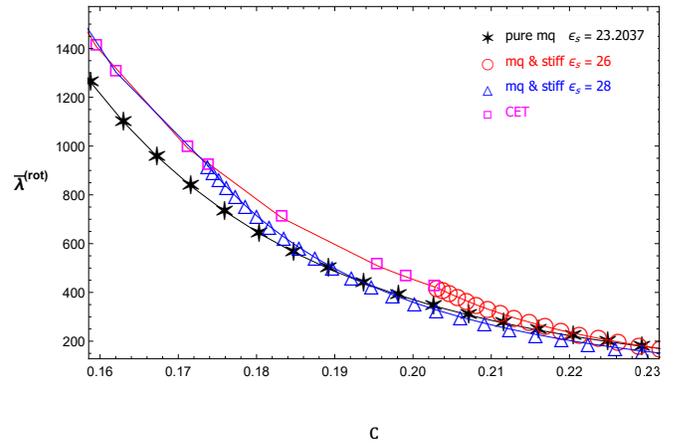}
	\caption{$\bar{\lambda}^{\rm (rot)}$ vs. $C$ around the transition region between NS with MQ core and CET NS }
	\label{fig:LCzoom}
\end{figure}

\begin{figure}[H]
	\centering
	\includegraphics[width=0.48\textwidth]{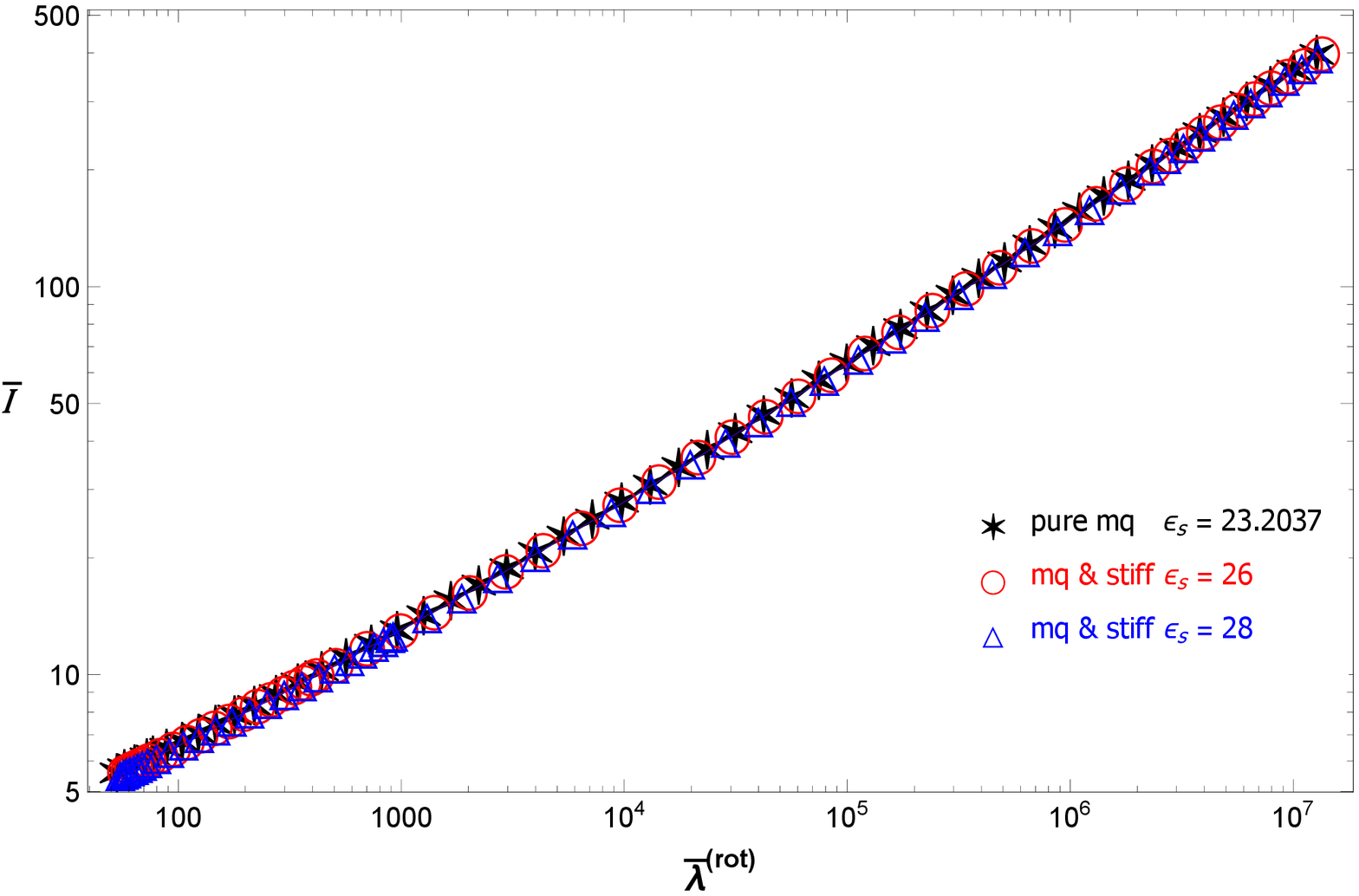}
	\includegraphics[width=0.48\textwidth]{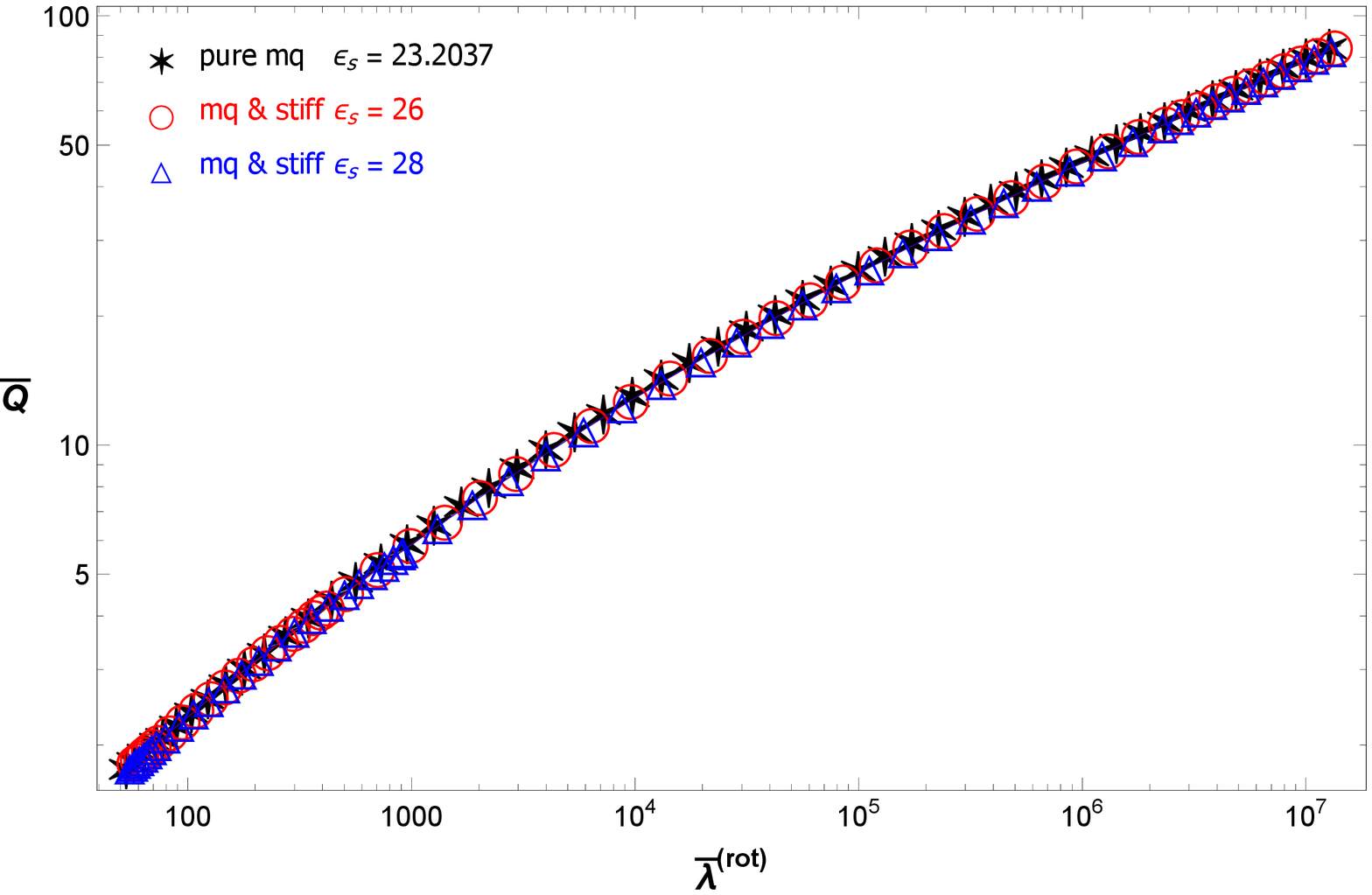}
	\caption{Upper: $\bar{I}$ vs. $\bar{\lambda}^{\rm (rot)}$ relation. Lower: $\bar{Q}$ vs. $\bar{\lambda}^{\rm (rot)}$ relation.}
	\label{fig:ILQ}
\end{figure}

\begin{figure}[H]
	\centering
	\includegraphics[width=0.48\textwidth]{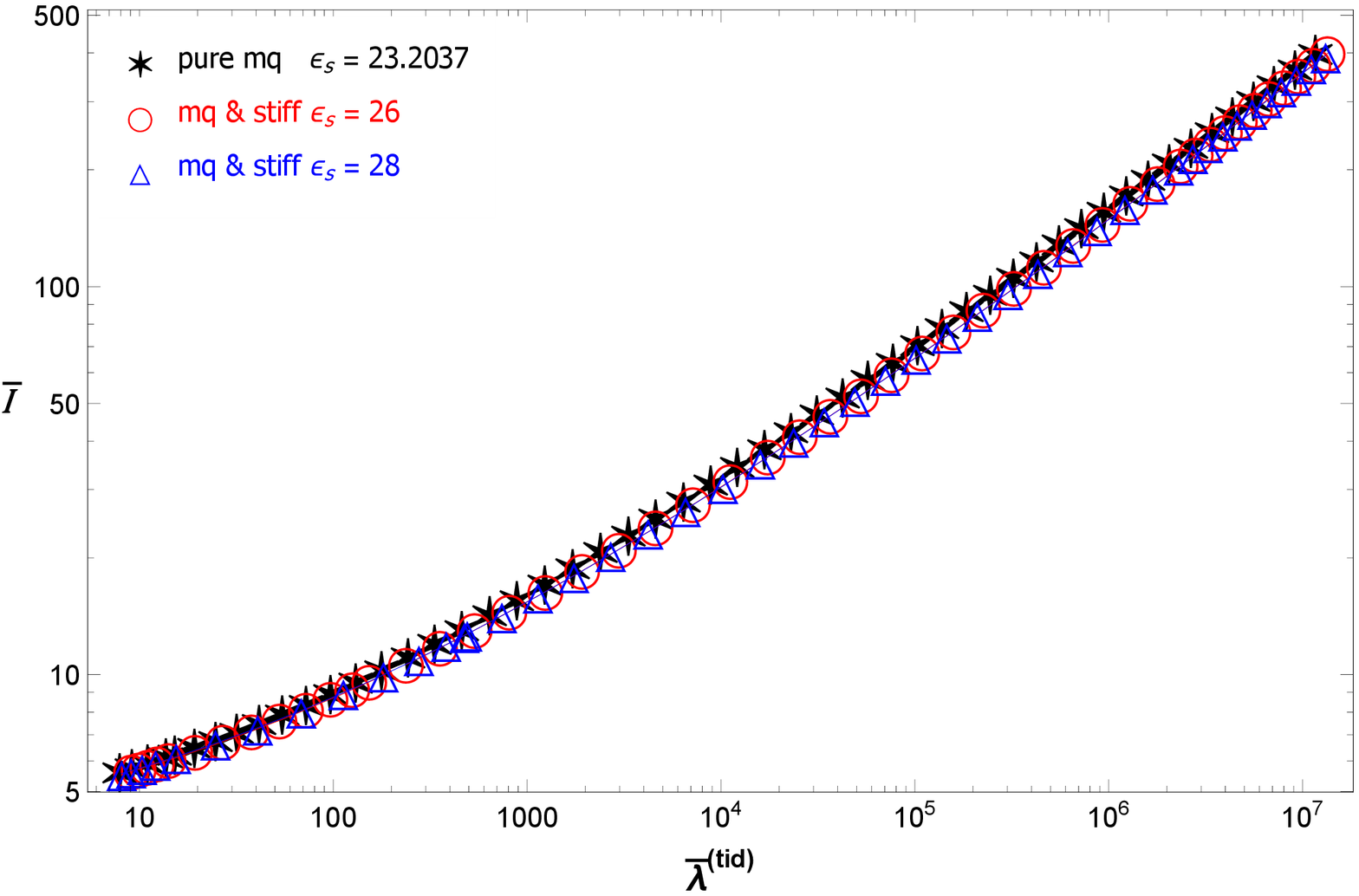}
	\includegraphics[width=0.48\textwidth]{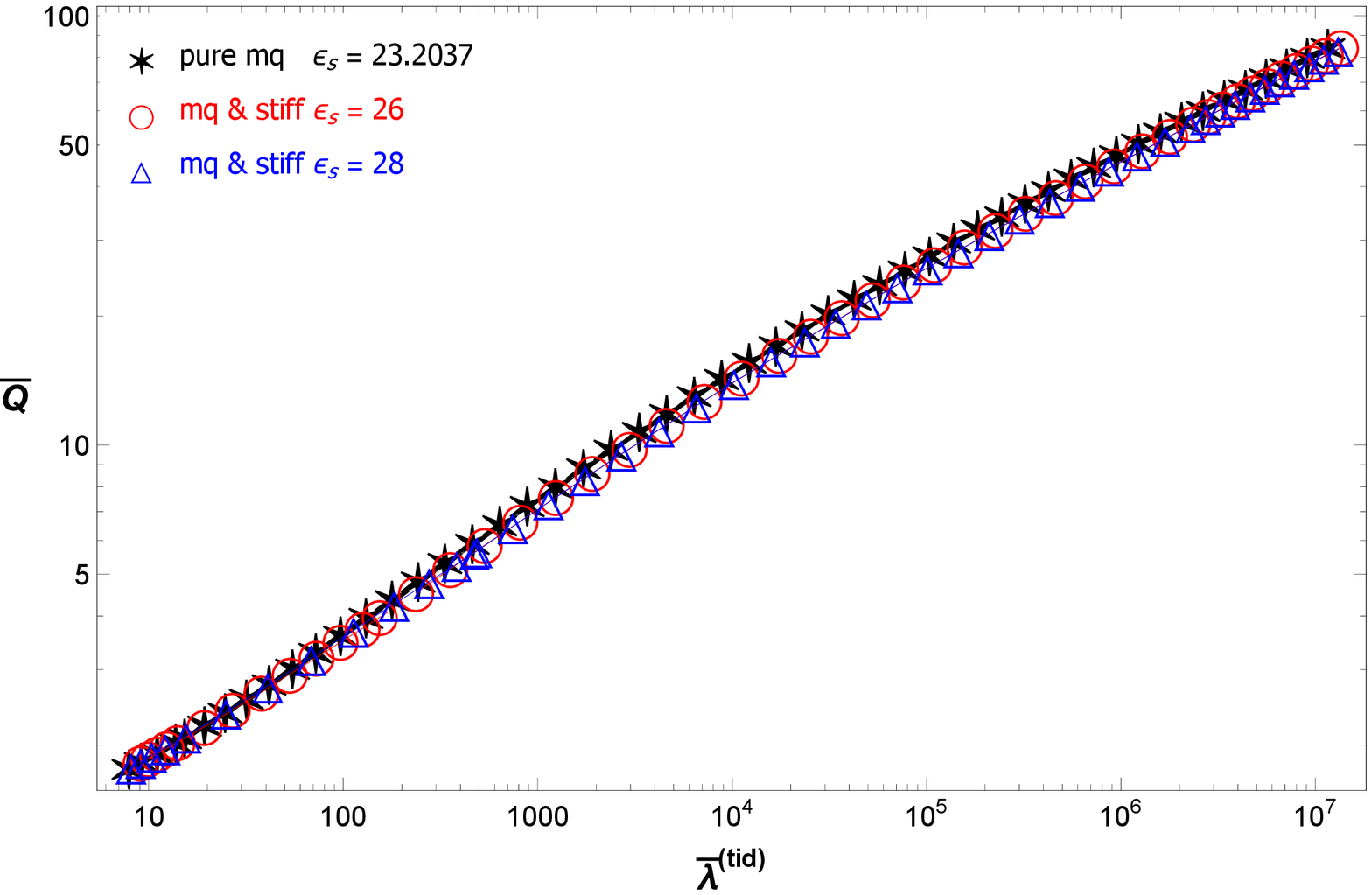}
	\includegraphics[width=0.48\textwidth]{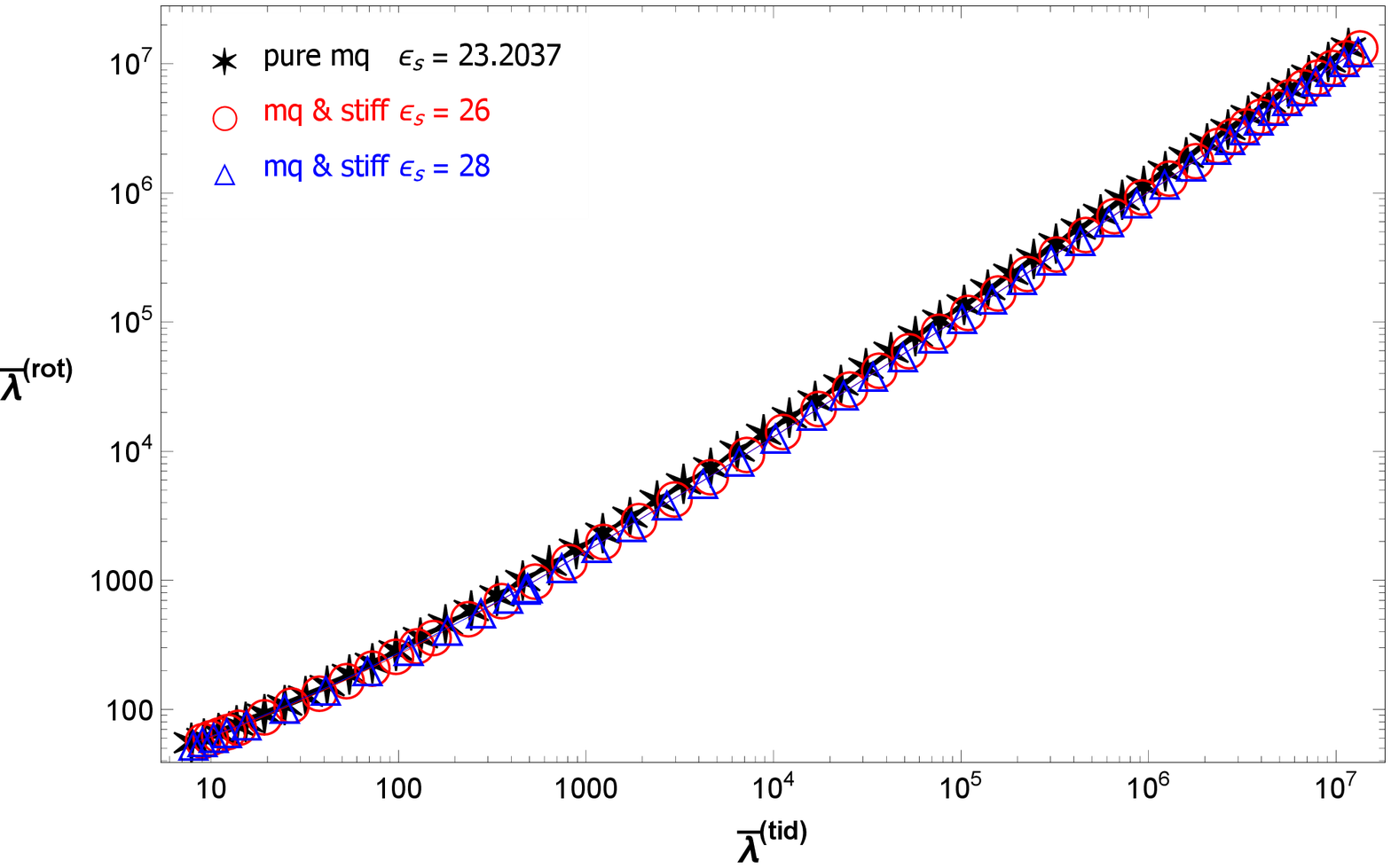}
	\caption{Upper: $\bar{I}$ vs. $\bar{\lambda}^{\rm (tid)}$ relation. Middle: $\bar{Q}$ vs. $\bar{\lambda}^{\rm (tid)}$ relation. Lower: $\bar{\lambda}^{\rm (rot)}$ vs. $\bar{\lambda}^{\rm (tid)}$ relation. }
	\label{fig:ILQtid}
\end{figure}

In this section, we shall display numerical results of $\bar{I}, \bar{\lambda}^{\rm (rot)}, \bar{Q}$ at linear and second order in spin of the NS with MQ core and pure MQ star.  The mass-radius diagram previously obtained in \cite{Pinkanjanarod:2020mgi} are shown in Fig.~\ref{fig:MR} for convenience in understanding the interior structure of the star.  As mentioned above, all quantities are independent of the rotation parameters in the perturbative regime of spin.  For completeness, the dimensionless tidal deformation parameter, $\bar{\lambda}^{\rm (tid)}$, calculated in Ref.~\cite{Pinkanjanarod:2021qto}~(denoted by $\Lambda$) is also plotted with $\bar{I}, \bar{\lambda}^{\rm (rot)}, \bar{Q}$ verifying the universal I-Love-Q relations involving both $\bar{\lambda}^{\rm (rot)}$ and $\bar{\lambda}^{\rm (tid)}$.

i.) $\bar{I}$ vs. $M, C$ in Fig.~\ref{fig:IMC}.  Near maximum mass, the NS with MQ core has distinct transition from MQ to CET in the value of moment of inertia.  This can be physically expected due to denser mass concentration in smaller region of the MQ phase resulting in smaller moment of inertia.  Since the MQ EoS in the low density regime is quite similar to the stiff CET EoS, the trend of the plots for lower masses is also similar, however not identical.  The difference can be seen in $\bar{I}$ vs. $C$ plot where at the same compactness, pure MQ star has larger $\bar{I}$ than the NS with MQ core. 

ii.) $\bar{Q}$ vs. $M, C$ in Fig.~\ref{fig:QMC}.  Again the plot $\bar{Q}$ vs. $M$ looks similar to $\bar{I}$ vs. $M$ and $\bar{\lambda}^{\rm (rot)}$ vs. $M$ except for the numerical values.  Transition between MQ core and CET crust is distinctive.  Differences between NS with MQ core and pure MQ star are manifest in $\bar{Q}$ vs. $C$ plot.

iii.) $\bar{\lambda}^{\rm (rot)}$ vs. $M, C$ in Fig.~\ref{fig:LMC}.  The relations between the rotational Love number with mass and compactness are interestingly similar to the moment of inertia parameter.  Highly dense and compact star results in small $\bar{\lambda}^{\rm (rot)}$ whereas transition between MQ core and nuclear crust can only be seen in the plot with mass of the star, and not in the plot with compactness $C$.  Difference between NS with MQ core and pure MQ, however, is distinctive in the compactness plot shown in Fig.~\ref{fig:LCzoom}. 

iv.) I-Love-Q~(rotation) relations in Fig.~\ref{fig:ILQ}.  Remarkable universality observed in \cite{Yagi:2013bca,Yagi:2013awa} of the relationship between $I, \lambda, Q$ can be seen from the plots between $\bar{I}$ vs. $\bar{\lambda}^{\rm (rot)}$ and $\bar{Q}$ vs. $\bar{\lambda}^{\rm (rot)}$.  This universal I-Love-Q relation thus can be used as a test for the validity of GR even though it cannot reveal the internal structure EoS of the NS unless the dependence on the mass and compactness are analysed.  

v.) I-Love-Q~(tidal) relations in Fig.~\ref{fig:ILQtid}.  As elaborated in details in \cite{Pinkanjanarod:2021qto}, the dimensionless deformation parameter $\bar{\lambda}^{\rm (tid)}$~(denoted by $\Lambda$ in \cite{Pinkanjanarod:2021qto}) can be calculated for the massive NS with MQ core.  Universality is also confirmed for $I, \bar{\lambda}^{\rm (tid)}, Q$ relations as well as ``Love-Love'' relation in the bottom of Fig.~\ref{fig:ILQtid}.

\section{Conclusions and Discussions}    \label{sec-con}

The multipole moments $I, \lambda^{\rm (rot, tid)}, Q$ generated by slow rotation of the massive NS with MQ core, CET NS and pure MQ are calculated using the EoS from the holographic SS model and stiff CET EoS for the nuclear crust.  Generically in the perturbative regime, the moment of inertia, rotational Love number and quadrupole moment are found to be determined purely by the zeroth-order star profile and independent of the rotation parameters.  Interestingly, the MQ core does not seem to violate the universal I-Love-Q relations found by Yagi and Yunes~\cite{Yagi:2013bca}.  However, analyses of $I, \lambda^{\rm (rot,tid)}, Q$ with respect to the mass and compactness of the star could reveal the existence of the MQ core or distinguish the hybrid star from the pure MQ star from the kink pattern as demonstrated in $\bar{I}, \bar{\lambda}^{\rm (rot, tid)}, \bar{Q}$ vs. $M, C$ plots.  Together with the kink in the MR diagram, the existence of NS with MQ core could be validated.

\begin{acknowledgments}

S.P.~(second author) is supported in part by the Second Century Fund: C2F PhD Scholarship, Chulalongkorn University. 	

\end{acknowledgments}

\end{document}